\documentclass[twocolumn,showpacs,preprintnumbers,prb]{revtex4}
\usepackage{amssymb}
\usepackage[dvips]{color}
\usepackage{epsfig}
\usepackage{dcolumn}
\usepackage{bm}

\def\G{\Gamma}

\def\D{\Delta}
\def\e{\epsilon}

\def\w{\omega}

\def\s{\sigma}

\begin{document}

\title{Spin-Dependent Ringing and Beats in a Quantum Dot System}
\author{Fabr{\'i}cio M. Souza}
\affiliation{International Centre for Condensed Matter Physics,
Universidade de Bras{\'i}lia, 70904-910, Bras{\'i}lia-DF, Brazil}
\pacs{PACS number}

\begin{abstract}
We report spin-dependent quantum coherent oscillations (ringing)
and beats of the total and the spin currents flowing through a quantum
dot with Zeeman split levels. The spin dependent transport is
calculated via nonequilibrium Green function in the transient
after a bias voltage is turned on at $t=0$. The dot is coupled to
two electrodes that can be ferromagnetic or nonmagnetic. In the
ferromagnetic case both parallel and antiparallel alignments are
considered. The coherent oscillation and beat frequencies are
controlled via the Zeeman energy $E_Z$. In particular, for $E_Z=0$
no beats are observed and the spin current is zero for nonmagnetic
leads. In the ferromagnetic case a finite spin current is found
for $E_Z=0$. The effects of temperature are also analyzed. We
observe that with increasing temperature the ringing response and
the beats tend to disappear. Additionally, the spin current goes
to zero for nonmagnetic leads, remaining finite in the
ferromagnetic case. The tunnel magnetoresistance (TMR) also
reveals quantum coherent oscillations and beats, and it attains
negative values for small enough temperatures and short times.
\end{abstract}

\volumeyear{year} \volumenumber{number} \issuenumber{number}
\eid{identifier}
\date[Date: ]{\today}
\maketitle

\section{Introduction}

Spin coherent dynamics and transport in quantum dots has attracted
a lot of attention due to its relevance to the potential new
generation of spintronic devices\cite{spintronics} (e.g.,
Datta-Das transistor\cite{dattadas} and storage
devices\cite{mef01,sc02,mk04}), and for quantum computation and
information processes,\cite{dpd95,dl98,man00} where quantum
coherence of the electron spin is desirable. Recent experiments
demonstrate the possibility to coherently manipulate quantum
states of single and double electron spins in quantum dot
systems.\cite{th03,jme04,jrp05,mvgd05,fhlk06,ag06,mvgd06,mhm07} Such
control can be achieved, for instance, via fast bias/gate voltage
pulses, electron spin resonance (ESR) fields and coherent optical
fields. Those experiments indicate the feasibility of using a
single electron in quantum dot system as a quantum bit and reveal
encouraging spin coherence lifetimes for quantum processing.

\begin{figure}[h]
\par
\begin{center}
\epsfig{file=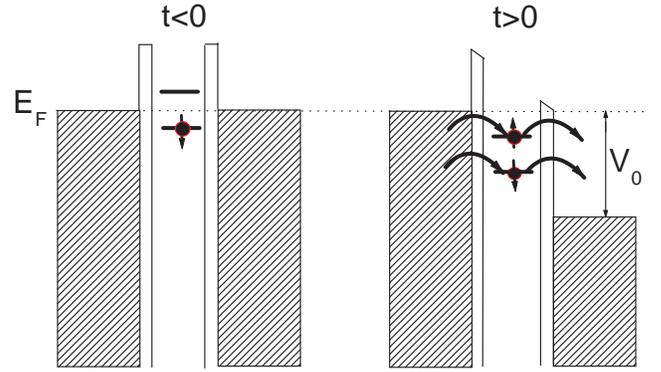, width=0.47\textwidth}
\end{center}
\caption{System studied: a quantum dot attached to two leads via
tunnel barriers. Both ferromagnetic and nonmagnetic leads are
considered. The dot level is Zeeman-split due to a external
magnetic field. Before the bias voltage is turned on ($t<0$) the
spin split level $\e_\downarrow$ is energetic accessible for the
electrons in the leads, while $\e_\uparrow$ ($>E_F$) is forbidden.
After a bias voltage $V_0$ is applied ($t>0$) both levels become
inside the conduction window and a spin-dependent tunnel current
arises.} \label{fig1}
\end{figure}

In addition, a variety of interesting spin-coherent effects have
been reported during recent years. For instance, coherent Rabi
oscillations generated by an ESR field were studied both
theoretically\cite{hae02} and experimentally\cite{fhlk06} in
quantum dot systems in the Coulomb blockade regime. Coherent
spin-dynamics was probed via time-resolved Faraday rotation (TRFR)
technique, revealing coherent dynamics of net carrier spins in
semiconductor bulk, quantum well and quantum
dots.\cite{sac97,jag99,jmk00} In addition to quantum coherent
oscillations, quantum beats of the spin magnetization due to
Zeeman splitting in a semiconductor quantum dot system were
observed via time-resolved Faraday rotation technique,\cite{jag99}
and quantum beats in the TRFR signal were measured for magnetic
fields higher than $5$T in a quantum dot system.\cite{ag06}
Quantum coherent beating in the magnetoresistance of a
double-barrier structure with both diluted magnetic semiconductor
(DMS) well and contacts was also predicted.\cite{jce01} Additionally to this, coherent 
oscillations of polarized current in magnetotransport through Zeeman split levels 
of quantum dots were investigated via quantum rate equations.\cite{sag05}

Quantum coherent oscillations (ringing) of the current flowing
through a single level quantum dot attached to nonmagnetic leads
was predicted in a transient time scale after a bias voltage is
turned on.\cite{nsw93,apj94,apj94_2,jm06} Here we describe
transport in the same transient regime but including
spin-dependent effects. An understanding of the spinfull case can
better guide possible experimental measurements of the ringing
response which now comes up with new spin-based signatures. Here
we apply the nonequilibrium Green function (NEGF) formulation
developed in [\onlinecite{apj94}] to account for (i) Zeeman
splitting of the dot level and/or (ii) ferromagnetic leads. These
two additional features introduce new spin-dependent effects. For
example, the coherent oscillations reported previously in
[\onlinecite{nsw93,apj94,apj94_2}] become spin-dependent with
slightly different frequencies for each spin component of the
current. This spin-splitting of the frequency generates quantum
coherent beats of the total current $I=I_\uparrow+I_\downarrow$
and of the spin current $I^s=I_\uparrow-I_\downarrow$. We also
find a dynamical tunnel magnetoresistance (TMR) that shows
coherent oscillations, coherent beats and attains negative values
in the transient regime for small enough temperatures.

The paper is organized as follows: in Sec. II we describe the
system and the formulation adopted, in Sec. III we present and
discuss the results and in Sec. IV we conclude.

\section{Model and Formulation}

Figure \ref{fig1} illustrates the system considered. It is
composed of one quantum dot coupled via tunnel barriers to a left
(L) and to a right (R) lead. Both ferromagnetic and nonmagnetic
leads are considered. For the ferromagnetic case both parallel (P)
and antiparallel (AP) alignments are analyzed. We assume that for
zero bias voltage ($t<0$) the spin-$\uparrow$ and
spin-$\downarrow$ levels are above and below the Fermi energy
$E_F$ of the reservoirs, respectively. This {\it prepared}
configuration gives rise to a relatively high transient spin
current in the case of nonmagnetic leads as we will describe in
Sec. III(a). When the bias voltage is turned on ($t>0$) both
levels attain resonance with the emitter states, thus resulting in
the subsequent spin dynamics.

The total Hamiltonian of the system is given by
$H=H_L+H_R+H_D+H_T$, where
\begin{equation}
H_{L/R}=\sum_{\mathbf{k} \s} \e_{\mathbf{k} \s L/R}(t)
c_{\mathbf{k} \s L/R}^\dagger c_{\mathbf{k} \s L/R},
\end{equation}
with $\e_{\mathbf{k} \s L/R}(t)=\e_{\mathbf{k} \s L/R}^0+
\Delta_{L/R} (t)$ being the free-electron energy with a
time-dependent contribution $\Delta_{L/R} (t)$ due to the bias
voltage. The operator $c_{\mathbf{k} \s L/R}$ ($c_{\mathbf{k} \s
L/R}^\dagger$) annihilates (creates) one electron with wave vector
$\mathbf{k}$ and spin $\s$ in the left $(L)$ or right $(R)$ lead.
For the dot Hamiltonian we have
\begin{equation}
H_D=\sum_{\s} [\e_d(t) + \s E_Z/2] d_\s^\dagger d_\s,
\end{equation}
where $\e_d(t)=\e_0+\Delta_d(t)$ is the dot level with a
time-dependent term $\Delta_d(t)$, $E_Z$ is the Zeeman energy due
to some external magnetic field\cite{rh03} and $\s=+$ or $-$ for
spin $\uparrow$ or $\downarrow$, respectively. We neglect the
Zeeman splitting of the leads $E_Z^{\mathrm{leads}}$ in our
model.\cite{Zeemanleads} The operators $d_\s$ and $d_\s^\dagger$
annihilate and create, respectively, one electron with spin $\s$
in the dot.\cite{previousresults} 
For the leads-dot coupling we have
\begin{equation}
H_T=\sum_{\mathbf{k} \s \eta} (V c_{\mathbf{k} \eta}^\dagger d_\s
+V^* d_\s^\dagger c_{\mathbf{k} \s \eta}),
\end{equation}
where $V$ is a constant coupling parameter. We do not account for
spin relaxation and spin decoherence in our model, which is
reasonable for short enough times. Previous studies have found for
the relaxation and decoherence times, $T_1$ and $T_2$
respectively, typically $T_1 \sim 1-20$
ms\cite{avk0001,jme04,mk04,joh} and $T_2 \gtrsim
\mu$s.\cite{jrp05,avk02,rs03}

The time dependent current is given by a sum of the currents
flowing into and out of the dot,
$I^\eta_\s(t)=I_\s^{\eta,in}(t)+I_\s^{\eta,out}(t)$. In the
noninteracting case and wideband limit they read\cite{hh96}
\begin{equation}\label{Iin}
I_\s^{\eta,in}(t)=-\frac{e\G_\s^\eta}{\hbar} \int \frac{d\e}{\pi}
f_\eta (\e) \mathrm{Im} [A_{\s \eta}(\e,t)],
\end{equation}
and
\begin{equation}\label{Iout}
I_\s^{\eta,out}(t)=-\frac{e\G_\s^\eta}{\hbar} \int
\frac{d\e}{2\pi} \sum_{\xi=L,R}\G_\s^\xi  f_\xi (\e) |A_{\s \xi}
(\e,t)|^2,
\end{equation}
where $\G_\s^\eta$ is the tunneling rate, $\G_\s^\eta=2\pi |V|^2
\rho_\s^\eta$, with $\rho_\s^\eta$ being the constant density of
states for spin $\s$ in lead $\eta$($=L,R$). The ferromagnetism of
the leads is accounted for via the model $\G_\s^L=\G_0(1 + \s p)$ and
$\G_\s^R=\G_0(1 \pm \s p)$, where $\G_0$ is the leads-dot coupling
strength and $p$ gives the polarization degree of the leads. The
signs $+/-$ in $\G_\s^R$ apply for parallel and antiparallel
configurations, respectively.\cite{wr01} The function $f_\eta
(\e)$ is the Fermi distribution function for lead $\eta$. For a
bias voltage of the kind $\D_L=0$, $\D_R=-V_0 \theta(t)$ and $\D_d=\D_R/2$
the function $A_{\s \eta} (\e,t)$ is given by\cite{first}
\begin{eqnarray}\label{A1}
&&A_{\s \eta} (\e,t>0)= \frac{e^{i[\e-\e_0-\s
E_Z/2-\D_d+\D_\eta+i(\G_\s^L+\G_\s^R)/2]t/\hbar}}{\e-\e_0-\s
E_Z/2+i(\G_\s^L+\G_\s^R)/2}+\nonumber\\&&\phantom{xxx}
\frac{1-e^{i[\e-\e_0-\s
E_Z/2-\D_d+\D_\eta+i(\G_\s^L+\G_\s^R)/2]t/\hbar}}{\e-\e_0-\s
E_Z/2-\D_d+\D_\eta+i(\G_\s^L+\G_\s^R)/2}.
\end{eqnarray}
Using this expression inside Eqs. (\ref{Iin}) and (\ref{Iout}) we
find the results presented in the next section. 

For comparison we apply the master equation (ME) technique to calculate the current and spin-current. While the NEGF is valid for both $k_BT \gg \G_0$ and $k_B T \ll \G_0$, the ME approach is accurate only for $k_BT \gg \G_0$ (sequential-tunneling limit). So we will restrict the comparison only for relatively high temperatures, which correspond to the range of mutual validity of the two approaches. The current expression in the ME formulation is given by\cite{me}
\begin{equation}\label{ImasterEq}
I_\s^\eta = e \G_\s^\eta [f_{\eta \s} P_0 - (1-f_{\eta \s}) P_\s +
\tilde{f}_{\eta \s} P_{\bar{\s}} - (1-\tilde{f}_{\eta \s}) P_2],
\end{equation}
where $P_0$, $P_\s$ and $P_2$ are the probabilities to have no
electron, one electron with spin $\s$ and two electrons,
respectively, in the dot. The Fermi functions are $f_{\eta
\s}=\{1+\mathrm{exp}[(\e_\s-\D_\eta)/(k_B T)]\}^{-1}$ and
$\tilde{f}_{\eta \s}=\{1+\mathrm{exp}[(\e_\s+U-\D_\eta)/(k_B
T)]\}^{-1}$ where $\e_\s=\e_d+\s E_Z / 2$. Defining the vector of
the occupation probabilities
$\mathbf{\underline{P}}=(P_0,P_\uparrow,P_\downarrow,P_2)^T$ we
can write the master equation as
\begin{equation}\label{Pdot}
\mathbf{\dot{\underline{P}}}=\mathbf{\underline{\underline{M}}}
\phantom{x} \mathbf{\underline{P}},
\end{equation}
where the transition matrix is given by
$\mathbf{\underline{\underline{M}}}=\mathbf{\underline{\underline{M}}^L}+\mathbf{\underline{\underline{M}}^R}$,
with
\begin{widetext}
\begin{equation}\label{MatrixM}
\mathbf{\underline{\underline{M}}^\eta}=\left(%
\begin{array}{cccc}
  -\G_\uparrow^\eta f_{\eta \uparrow}-\G_\downarrow^\eta f_{\eta \downarrow} & \G_\uparrow^\eta(1-f_{\eta\uparrow}) & \G_\downarrow^\eta(1-f_{\eta\downarrow}) & 0 \\
  \G_\uparrow^\eta f_{\eta\uparrow} & -\G_\uparrow^\eta(1-f_{\eta\uparrow})-\G_\downarrow^\eta \tilde{f}_{\eta\downarrow} & 0 & \G_\downarrow^\eta(1-\tilde{f}_{\eta\downarrow})\\
  \G_\downarrow^\eta f_{\eta\downarrow} & 0 & -\G_\uparrow^\eta \tilde{f}_{\eta\uparrow}-\G_\downarrow^\eta (1-f_{\eta\downarrow}) & \G_\uparrow^\eta(1-\tilde{f}_{\eta\uparrow})\\
  0 & \G_\downarrow^\eta\tilde{f}_{\eta\downarrow} & \G_\uparrow^\eta \tilde{f}_{\eta\uparrow} & \G_\uparrow^\eta \tilde{f}_{\eta\uparrow}+\G_\downarrow^\eta \tilde{f}_{\eta\downarrow}-(\G_\uparrow^\eta+\G_\downarrow^\eta) \\
\end{array}%
\right).
\end{equation}
\end{widetext}
If we take $f_{R \s}=\tilde{f}_{R \s}=0$ we obtain a matrix form
for $\mathbf{\underline{\underline{M}}}$ similar to the one
presented in [\onlinecite{gk03}]. Even though Eqs.
(\ref{ImasterEq})-(\ref{MatrixM}) can account for Coulomb
interaction in the sequential limit, we are interested in comparing
the results obtained from both the ME and the NEGF (which is
noninteracting in the present formulation), thus we simply assume
$U=0$ in the above ME expressions. Some effects of Coulomb
interaction in a transient response (sequential limit) can be
found, for instance, in [\onlinecite{fms07}].

\section{Results}

\subsection{Nonmagnetic Leads}

\begin{figure}[h]
\par
\begin{center}
\epsfig{file=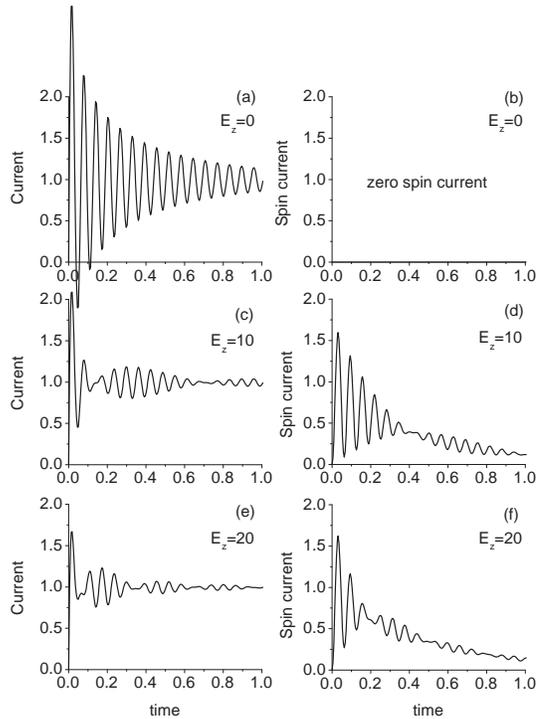, width=0.39\textwidth}
\end{center}
\caption{Total current $I_\uparrow+I_\downarrow$ (left panels) and
spin current $I_\uparrow-I_\downarrow$ (right panels) against time
for different Zeeman energies $E_Z$. For $E_Z=0$ the current shows
coherent oscillations (ringing) with a period $T_c=2\pi \hbar/|
E_F-\e_d |$. No spin current is observed for $E_Z=0$. In contrast,
for $E_Z=10$ and $E_Z=20$ a spin current arises and quantum
coherent beats are seen in both current and spin current. Units:
$e\G_0/\hbar$ for the currents, $\hbar/\G_0$ for the time and
$\G_0$ for the energies. Parameters: $\e_0=0$, $V_0=200$, $k_B
T=0.1$, $p=0$.} \label{fig2}
\end{figure}

Figure \ref{fig2} shows the time evolution of the total current
$I=I_\uparrow+I_\downarrow$ and the spin current
$I^s=I_\uparrow-I_\downarrow$ in the emitter
lead,\cite{commentemitter} obtained via NEGF. Different Zeeman
energies are considered. For $E_Z=0$ there is no spin current
[Fig. \ref{fig2}(b)] and the current [Fig. \ref{fig2}(a)] presents
the typical coherent oscillations that arise after a bias voltage
is turned on.\cite{nsw93,apj94,apj94_2,jm06} The oscillatory
frequency is given by $\hbar \w_c=|E_F-\e_d|$ (or equivalently the period $T_c=2\pi \hbar
/|E_F-\e_d|$) and the damping is due to the leads-dot coupling. We
may note that for the parameters adopted in Fig. \ref{fig2}(a) we
have $T_c \approx 0.06 \hbar/\G_0$.\cite{typical} In the presence
of Zeeman splitting the frequency $\w_c$ becomes spin dependent,
with $\hbar \w_c^\s=|E_F-\e_d-\s E_Z/2|$, so the spin components
of the current oscillate with distinct frequencies. In the case of
relatively close frequencies, quantum beats of $I$ and $I^s$ are
seen, with beating frequency given by $\hbar|\w_c^\uparrow -
\w_c^\downarrow|= E_Z$ [Figs. \ref{fig2}(c)-(f)].

The spin current observed in the presence of Zeeman energy [Figs.
\ref{fig2}(d) and (f)] comes from the initial configuration
($t<0$) of the system. For $t<0$ the occupations are $n_\downarrow
\approx 1$ and $n_\uparrow \approx 0$. When the bias voltage is
turned on at $t=0$ the spin $\uparrow$ electrons in the emitter
lead can start to flow through the dot, while the spin
$\downarrow$ ones have to wait till the initial $\downarrow$
electron in the dot leaves to the collector lead (Pauli blockade).
Since this tunnel event takes a typical time of $\hbar/\G_0$ a
spin current is expected in this time range. In the case of
initially both $n_\uparrow \approx 0$ and $n_\downarrow \approx
0$, which can be achieved by taking $\e_0$ high enough in order to
forbid thermal occupation of the spin split levels, an oscillatory
spin current around zero is observed.

\begin{figure}[tbp]
\par
\begin{center}
\epsfig{file=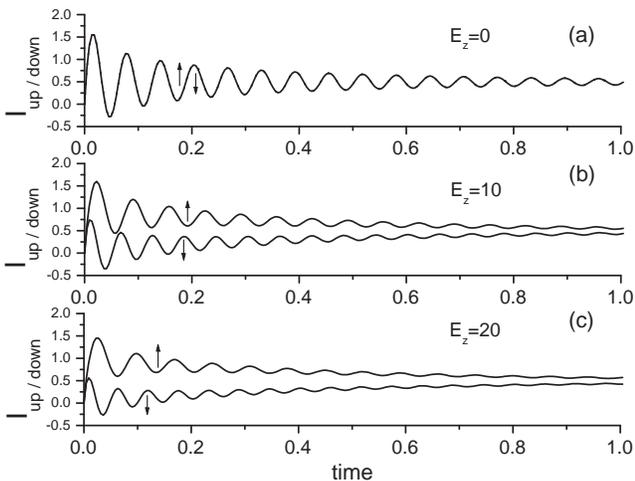, width=0.47\textwidth}
\end{center}
\caption{Spin resolved currents (denoted by up and down arrows
close to the curves) against time for three different Zeeman
energies. For $E_Z=0$, $I_\uparrow$ and $I_\downarrow$ coincide.
As $E_Z$ increases we find a red-shift for the frequency of
$I_\uparrow$ and a blue-shift for the frequency of $I_\downarrow$.
This frequency splitting results in the quantum beats seen in Fig.
\ref{fig2}. Units and parameters as in Fig. \ref{fig2}.}
\label{fig3}
\end{figure}

In Fig. (\ref{fig3}) we show $I_\uparrow$ and $I_\downarrow$
separately. For $E_Z=0$ the currents $I_\uparrow$ and
$I_\downarrow$ coincide, while for $E_Z=10$ and $E_Z=20$ they
differ, thus generating spin current. Eventually for high enough
times ($t \gg 1$) both $I_\uparrow$ and $I_\downarrow$ become close
to each other, thus resulting in $I^s \rightarrow 0$. The
difference in frequency for each spin component is clearly seen in
Figs. \ref{fig3}(b)-(c). In particular as $E_Z$ increases, we
observe a blue shift for the frequency $\w_c^\downarrow$ and a red
shift for $\w_c^\uparrow$.

\begin{figure}[tbp]
\par
\begin{center}
\epsfig{file=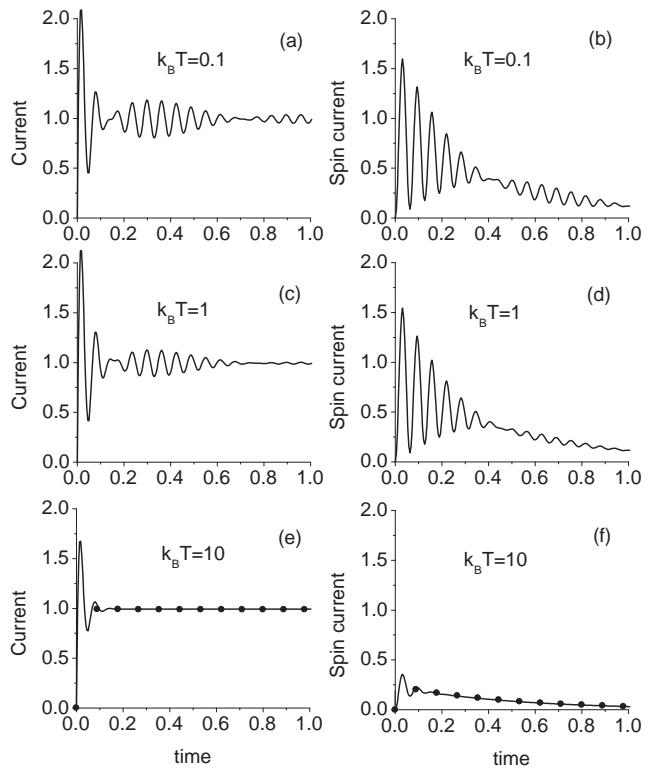, width=0.47\textwidth}
\end{center}
\caption{Total current (left panels) and spin current (right
panels) against time for differing temperatures. Quantum coherent
oscillations and beats are seen for $k_B T=0.1$ and $k_B T=1$. As
the temperature increases the oscillations vanish (see $k_B
T=10$). Additionally, the spin current is significantly suppressed
for $k_B T=10$. For comparison in panels (e) and (f) we show the
results obtained via master equation (dots). Units as in Fig.
\ref{fig2}. Parameters: $\e_0=0$, $V_0=200$, $E_Z=10$, $p=0$.}
\label{fig4}
\end{figure}

In Fig. (\ref{fig4}) we analyze the effects of temperature on the
coherent oscillations. We find that as $k_B T$ increases the
oscillatory behavior tends to disappear in both the current and
the spin-current. This is expected since with increasing $T$ the
coherence is washed out. For comparison we show in Fig.
\ref{fig4}(e)-(f) the total and the spin currents calculated via
the master equation technique (dots). We find close agreement
between the nonequilibrium Green function and the master equation
results, except by the strongly suppressed coherent oscillations still seen
in the NEGF calculation for $k_B T=10$. As the temperature increases these coherent 
oscillations (ringing) disappear and both NEGF and ME approaches give equal results. 
Additionally to this, we note that the spin current is suppressed with increasing $k_B T$.
For even higher temperatures we find zero spin current for all
times, which is related to the fact that $n_\uparrow(t<0)$
approaches $n_\downarrow (t<0)$ with increasing $k_B T$--thus
resulting in a less effective transient Pauli blockade. When the leads are
ferromagnetic, though, the source of spin current is not only the
initial configuration of the system but also the leads itself, so
$I^s$ remains nonzero even for high temperatures, as we describe
next.

\subsection{Ferromagnetic Leads}

\begin{figure}[tbp]
\par
\begin{center}
\epsfig{file=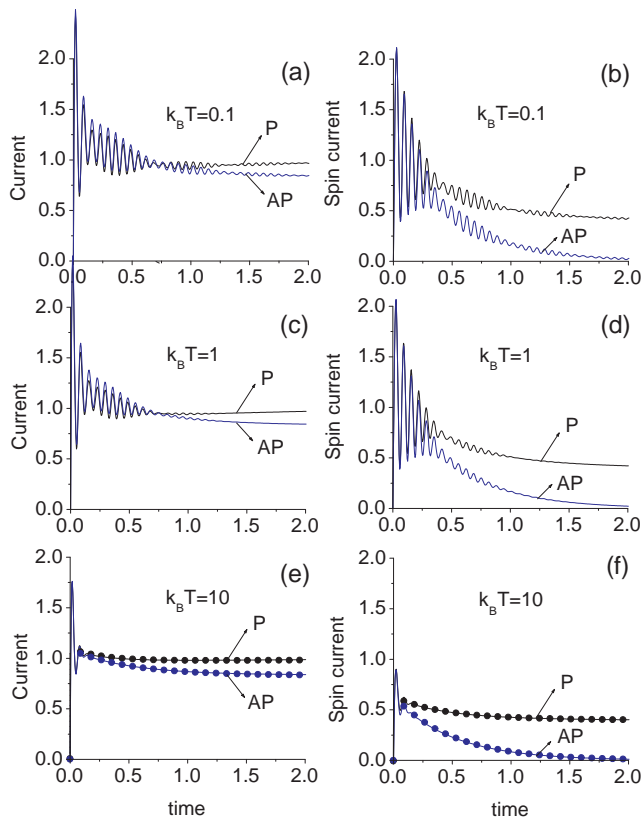, width=0.47\textwidth}
\end{center}
\caption{(Color online) Total current $I=I_\uparrow+I_\downarrow$ (left panels)
and spin current $I^s=I_\uparrow-I_\downarrow$ (right panels)
against time for differing temperatures in the case of
ferromagnetic leads. Both parallel (P) and antiparallel (AP) configurations
are shown. Coherent oscillations and beats are seen in both
alignments. The standard inequality $I_P>I_{AP}$
(magnetoresistance effect) is seen for $t \gtrsim 1$ and $k_B
T=0.1$ or $k_BT=1$. For times $t \lesssim 1$ unusual behaviors
like $I_P<I_{AP}$ can be found for these temperatures. In
contrast, for $k_B T=10$ almost no oscillations are seen and
$I_P>I_{AP}$ is recovered for all times. For comparison in panels
(e)-(f) we show results obtained via master equation technique
(dots). We also note that the spin current goes to zero as the
time evolves in the AP alignment while remains finite in the P
case even for $k_B T=10$. Units as in Fig. \ref{fig2}. Parameters:
$\e_0=0$, $V_0=200$, $E_Z=10$, $p=0.4$.} \label{fig5}
\end{figure}

Figure (\ref{fig5}) shows the current and the spin current against
time in the presence of ferromagnetic leads for differing
temperatures and $E_Z=10$. Both P and AP configurations are
analyzed. Here we have enlarged the time scale compared to Fig.
\ref{fig4} because some contrasting features between P and AP
alignments are better seen after $t=1$ (e.g., $I_P
> I_{AP}$). For $t \lesssim 1$ the standard inequality $I_P
> I_{AP}$ is only true when we approach the sequential limit for $k_B T=10$.
Curiously, for $k_B T=0.1$ and $k_B T=1$ we find $I_{AP}$ slightly
above $I_P$ for $t \lesssim 1$. This unusual behavior can be
qualitatively understood in terms of tunnel rates and the initial (\emph{prepared})
spin-population of the dot. Initially ($t<0$) the dot is occupied
by one spin $\downarrow$ electron. When the bias voltage is
applied ($t>0$) this electron can leave the dot to the collector
lead with a tunnel rate given by $\G_\downarrow^R$. Since
$\G_\downarrow^R$ is bigger for AP than for P configuration, the
initial spin $\downarrow$ electron leaves the dot faster in the AP
alignment. This turns the Pauli blockade less effective in this
configuration, thus allowing the emitter current $I_{AP}$ to be
slightly greater than $I_P$ at initial times.

When $k_B T$ exceeds the coupling strength $\G_0$, the coherent
oscillations of both the current and the spin current are strongly
suppressed and the standard inequality $I_P>I_{AP}$ is recovered
even for short times after $t=0$. Here we also see an agreement
between the results obtained via NEGF and ME, except by the coherent oscillations residually present in the NEGF calculation for $k_B T=10$. 
As the temperature increases even further these residual 
oscillations of the current and the spin-current are washed out and both NEGF and ME 
approaches converge to the same results. Finally, we observe that the spin current does not
tend to zero with increasing temperature as we see in the
nonmagnetic case [compare Figs. \ref{fig4}(f) and \ref{fig5}(f)].
The suppression of $I^s$ is observed only with increasing time in
the AP alignment, remaining finite for the stationary limit in the
P configuration.

\begin{figure}[t]
\par
\begin{center}
\epsfig{file=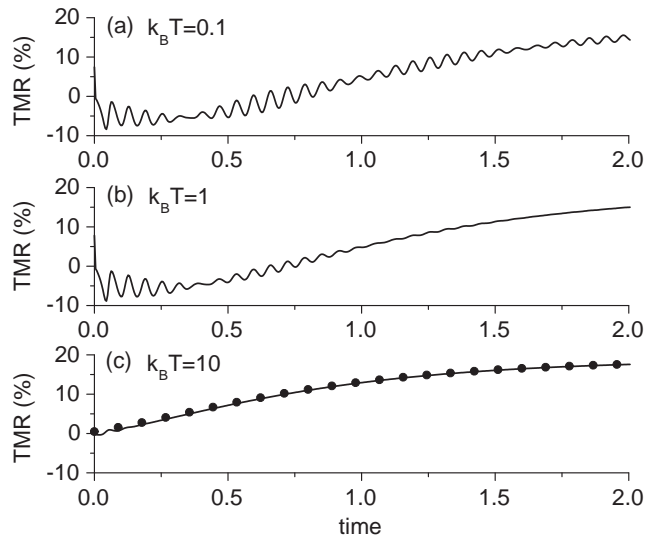, width=0.47\textwidth}
\end{center}
\caption{Tunnel magnetoresistance (TMR) against time for
increasing $k_B T$. For $k_B T=0.1$ and 1 the TMR shows a ringing
response, quantum beats and attains negative values before it
increases to a stationary value close to $20\%$. In contrast, for
$k_B T=10$ the TMR is always positive with almost no oscillations.
The dots show the TMR obtained via the master equation technique.
Units as in Fig. \ref{fig2}. Parameters: $\e_0=0$, $V_0=200$,
$E_Z=10$, $p=0.4$.} \label{figtmr}
\end{figure}

In Fig. (\ref{figtmr}) we show the time-dependent TMR, defined as
TMR=$(I_P-I_{AP})/I_{AP}$. Coherent oscillations and quantum beats
are seen for both $k_B T=0.1$ and $1$ and is only residually
observed for $k_B T=10$. Interestingly, the TMR attains negative
values for short times and small temperatures.\cite{comment3} In
panel (c) we compare the TMR obtained via NEGF (solid line) and ME
(dots) with essentially no contrast between them.

\begin{figure}[t]
\par
\begin{center}
\epsfig{file=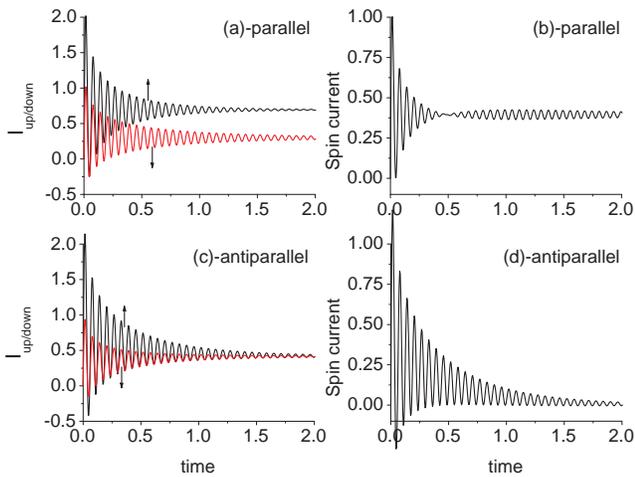, width=0.47\textwidth}
\end{center}
\caption{(Color online) Spin resolved currents (denoted by up and down arrows
closed to the curves) and spin current for both P and AP
alignments and $E_Z=0$. Both $I_\uparrow$ and $I_\downarrow$
oscillate with the same frequency due to $E_Z=0$, but in contrast
to the nonmagnetic case [Fig. \ref{fig3}(a)] here the amplitude
and decaying rate ($\G_\s^L+\G_\s^R$) can differ for each spin
component. In particular in the P alignment the interplay between
the amplitudes and the decaying rates gives rise to a node in the
spin current. Units as in Fig. \ref{fig2}. Parameters: $\e_0=0$,
$V_0=200$, $E_Z=0$, $k_B T=0.1$, $p=0.4$.} \label{fig7}
\end{figure}

Finally in Fig. \ref{fig7} we show the spin resolved currents and
the spin current in the ferromagnetic case for $E_Z=0$. Here $I^s
\neq 0$ [in contrast to the nonmagnetic case, Fig. \ref{fig2}(b)] and no beats are
seen since $I_\uparrow$ and $I_\downarrow$ have the same
frequency. Additionally, the oscillation amplitudes and the
decaying rates can differ for each spin component. In the parallel
configuration we observe that $I_\uparrow$ starts oscillating with
a higher amplitude than $I_\downarrow$ but its oscillations are
faster suppressed. This is due to the inequality
$\G_\uparrow^L+\G_\uparrow^R > \G_\downarrow^L +\G_\downarrow^R$.
The interplay between oscillation amplitudes and decaying rates
gives rise to a \emph{node} in the spin current seen around
$t=0.5$ [Fig. \ref{fig7}(b)]. For long enough times both
$I_\uparrow$ and $I_\downarrow$ attain their respective stationary
values with $I_\uparrow > I_\downarrow$ [Fig. \ref{fig7}(a)]. In
the AP alignment $I_\uparrow$ starts oscillating with a higher
amplitude than $I_\downarrow$ but, in contrast to the P case, the
decaying rate is the same for both spin component
($\G_\uparrow^L+\G_\uparrow^R=\G_\downarrow^L+\G_\downarrow^R$).
Consequently no node is observed in the spin current, which simply
oscillates decaying to zero.

\section{Conclusion}

We have calculated spin-dependent transport through a Zeeman split
quantum dot level after a bias voltage is turned on. The dot is
coupled to two leads (source and drain) that can be ferromagnetic
or nonmagnetic. Quantum coherent oscillations (ringing) and beats
of the current and of the spin current are found in a transient
time scale. The frequency of the beats can be tuned via the Zeeman
splitting energy $E_Z$. We also report a spin current $I^s$ that
arises when $E_Z \neq 0$ and $k_B T < \G_0$ for nonmagnetic leads.
For the ferromagnetic case the spin current can be seen even for
$E_Z=0$ and high temperatures ($k_B T >\G_0$). In particular $I^s$
goes to zero as the time evolves in the AP alignment and remains
finite in the stationary limit for the P configuration. The tunnel
magnetoresistance was also analyzed. We found quantum coherent
beats and negative values of the TMR due to the transient dynamics
of incoming and outgoing spin polarized electrons in the quantum
dot. The negative vales of the TMR vanish as the temperature
increases, which is confirmed by master equation calculations.

The author acknowledges A. P. Jauho, J. C. Egues and J. P. Morten for valuable
comments and suggestions. The author also acknowledges the kind
hospitality at MIC-DTU (Denmark) during the final stages of this
work. This work was supported by the Brazilian Ministry of Science and
Technology and IBEM (Brazil).


\begin{thebibliography}{99}
\bibitem{Zeemanleads} In principle this is experimentally possible by applying a
local magnetic field or by using a lead's material with small
enough $g$ factor in order to have $E_Z^{\mathrm{leads}} \ll E_F$.
\bibitem{first} This expression was originally derived in
[\onlinecite{nsw93}] for the spin independent case.
\bibitem{commentemitter} Here we present the current and the spin current in the
emitter lead. The collector current and spin current show no
significant contrasting features.
\bibitem{comment3} The negative values of the TMR observed here are not physically
equivalent to those seen in [\onlinecite{fms07}], where negative
values are found for $k_B T \gg \G_0$.
\end{thebibliography}
\end{document}